\documentclass[12pt,preprint]{aastex}

\newcommand{\HI}{\textsc{H\,i}}

\shorttitle{Disk-halo magnetic field transport in NGC~6946}
\shortauthors{Heald}

\begin{document}

\title{Magnetic field transport from disk to halo via\\the galactic chimney process in NGC~6946}

\author{George H. Heald\altaffilmark{1}}
\affil{ASTRON, Postbus 2, 7990 AA Dwingeloo, The Netherlands}
\email{heald@astron.nl}

\begin{abstract}
The interstellar medium (ISM) in galaxies is directly affected by the mass and energy outflows originating in regions of star formation. Magnetic fields are an essential ingredient of the ISM, but their connection to the gaseous medium and its evolution remains poorly understood. Here we present the detection of a gradient in Faraday rotation measure (RM), co-located with a hole in the neutral hydrogen (\HI) distribution in the disk of the nearby spiral galaxy NGC~6946. The gas kinematics in the same location show evidence for infall of cold gas. The combined characteristics of this feature point to a substantial vertical displacement of the initially plane-parallel ordered magnetic field, driven by a localized star formation event. This reveals how the large-scale magnetic field pattern in galaxy disks is directly influenced by internal energetic phenomena. Conversely, magnetic fields are observed to be an important ingredient in disk-halo interactions, as predicted in MHD simulations. Turbulent magnetic fields at smaller spatial scales than the observed RM gradient will also be carried from the disk and provide a mechanism for the dynamo process to amplify the ordered magnetic field without quenching. We discuss the observational biases, and suggest that this is a common feature of star forming galaxies with active disk-halo flows.
\end{abstract}


\section{Introduction}

The disk-halo interface is crucial for understanding the evolution of spiral galaxies. Star formation drives vertical hot gas flows in a process which is described by the galactic fountain \citep{shapiro_field_1976,bregman_1980} or chimney model \citep{norman_ikeuchi_1989}. In both models, the outflowing hot gas eventually condenses into cool clouds that fall back down onto the disk on timescales of tens of Myr. These returning clouds are thought to make up at least a part of the population of high and intermediate velocity clouds (HVCs and IVCs). The returning cool gas flow must be interacting with the ambient disk-halo medium, as is indicated by the HVC and IVC morphology \citep{bruens_etal_2000,heitsch_putman_2009} as well as the kinematics of thick disks in external galaxies and the Milky Way \citep{heald_etal_2007,fraternali_binney_2008,kalberla_kerp_2009}. Magnetic fields appear to be essential in reproducing the structure and evolution of the multiphase ISM \citep[e.g.,][]{avillez_breitschwerdt_2005,hill_etal_2012} and the longevity of infalling clouds \citep{santillan_etal_2004}.

NGC~6946 is a nearby \citep[$D=6.8\,\mathrm{Mpc}$;][]{karachentsev_etal_2000} grand-design spiral galaxy that has been studied at a broad range of wavelengths. It has a substantial integrated star formation rate \citep[$2.8\,M_\odot\,\mathrm{yr}^{-1}$;][]{calzetti_etal_2010}, together with a large number of catalogued holes in the \HI\ density distribution \citep{boomsma_etal_2008} that have sizes, masses, and energetics consistent with formation by localized pockets of star formation activity. For many of these features, the stars which provided the required energy are not distinctly visible, but this is typical of \HI\ holes \citep[e.g.,][]{brinks_bajaja_1986,bagetakos_etal_2011,warren_etal_2011}. The properties of the magnetic field ($\vec{B}$) in NGC~6946 have also been studied in detail, showing that its energy density is comparable to other ISM components, and perhaps even dominates at large galactocentric radii \citep{beck_2007}. The large-scale magnetic field has been probed using polarized synchrotron radiation and its Faraday rotation measure ($\mathrm{RM}\propto\int\,n_e\vec{B}\cdot\mathrm{d}\vec{l}$ where $n_e$ is the thermal electron density and $\vec{l}$ is the line of sight), both of which trace the ordered field. Faraday RM is a sensitive tracer of the line of sight magnetic field, and is robustly measured using the RM Synthesis technique \citep{brentjens_debruyn_2005,heald_etal_2009}. As in other spirals, NGC~6946 has an axisymmetric spiral pattern in the disk combined with a quadrupolar poloidal component that becomes dominant at large vertical distances from the disk \citep{braun_etal_2010}. High-quality \HI\ line data \citep{boomsma_etal_2008} and radio continuum polarization data from a combined dataset spanning $1300-1432$ and $1631-1763$ MHz \citep{heald_etal_2009} are available in the literature.

\section{A magnetized \HI\ bubble}

\begin{figure}
\centering
\includegraphics[width=0.9\textwidth]{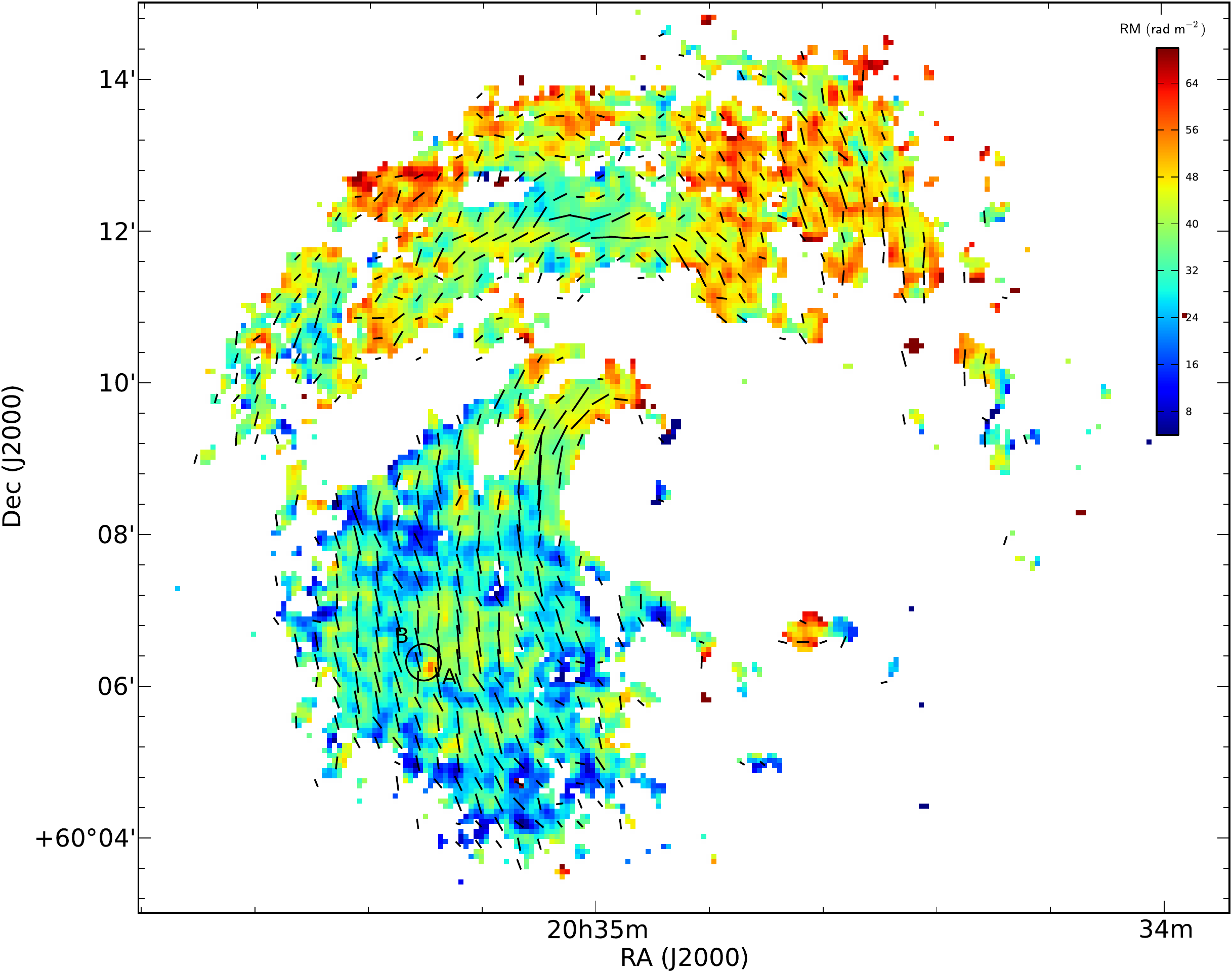}
\caption{RM image of NGC~6946 from \citet{heald_etal_2009}. Magnetic field vectors are shown with lines. The \HI\ hole catalogued by \citet{boomsma_etal_2008} and found to be colocated with the RM gradient is indicated with a black ellipse.}
\label{fig:rmhole}
\end{figure}

\begin{figure*}
\centering
\includegraphics[width=0.95\textwidth]{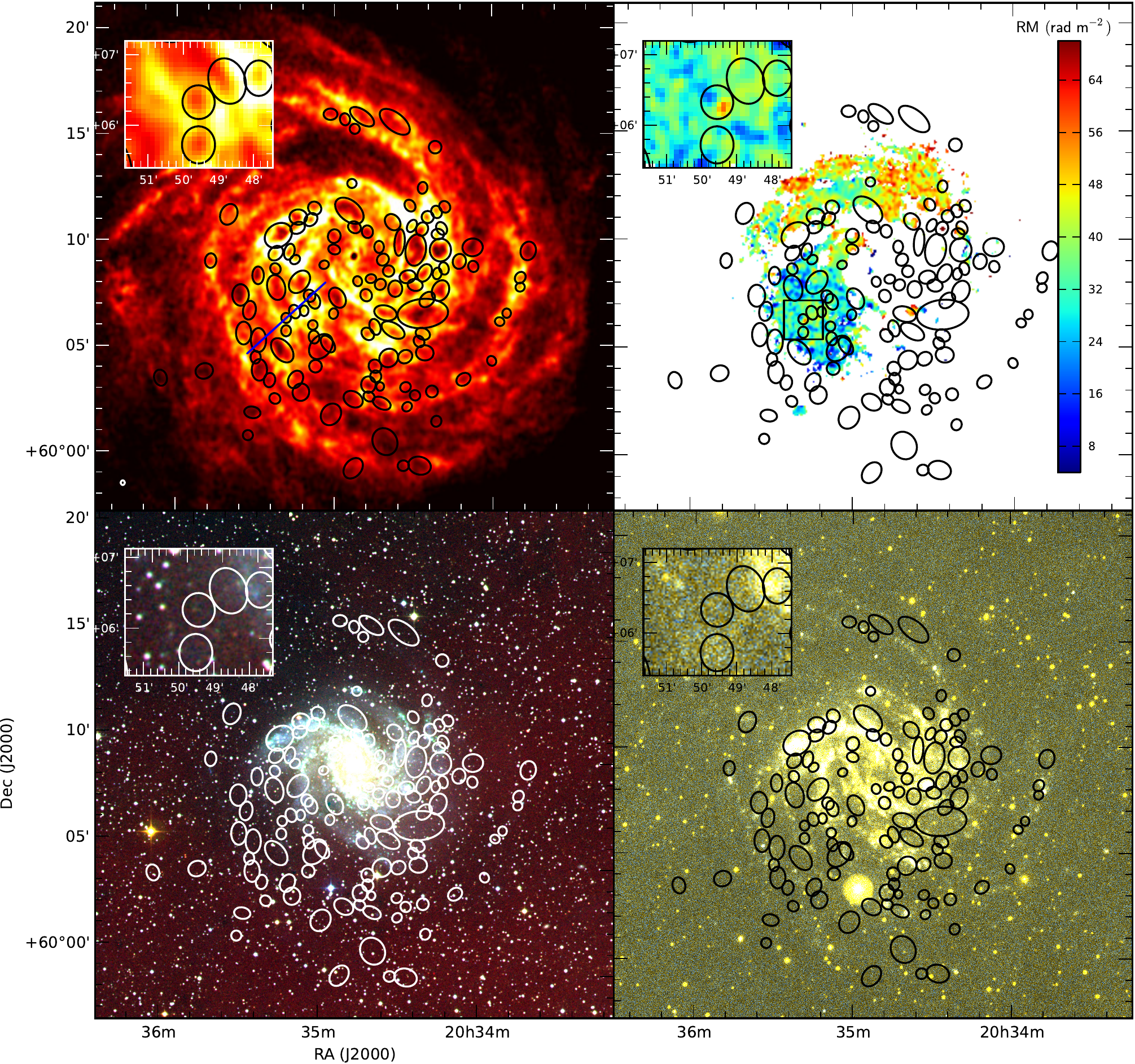}
\caption{Images of NGC~6946. All panels are presented on the same angular scale and indicate the locations of the \HI\ holes catalogued by \citet{boomsma_etal_2008} with black ellipses. Insets show the immediate vicinity of the feature described in the text. Top left: \HI\ column density. The blue line shows the slice used to create the PV diagram in Figure~\ref{fig:pv}. Top right: RM image. Bottom left: optical image (created from DSS-II $B$,$R$,IR plates). Bottom-right: GALEX Nearby Galaxy Survey \citep[NGS;][]{gildepaz_etal_2007} FUV-NUV false color image.}
\label{fig:grid}
\end{figure*}

The image of RM \citep{heald_etal_2009} corresponding to the widespread diffuse polarized synchrotron radiation in the disk of NGC~6946 is displayed in Figure~\ref{fig:rmhole}, and reveals a remarkable feature at $(\alpha,\delta)_{\mathrm{J2000.0}}\,=\,(20^h35^m18^s.2,60^\circ06^\prime18^{\prime\prime})$. There, a clear RM gradient is co-located with one of the \HI\ holes (see also Figure~\ref{fig:grid}) previously identified in the literature \citep{boomsma_etal_2008}. The RM at that location ranges from $18.8-57.0\,\mathrm{rad\,m^{-2}}$. The midpoint of the RM gradient is close to the typical RM value in the surrounding area of the disk, about $40\,\mathrm{rad\,m^{-2}}$. The RM gradient is on a scale of $17^{\prime\prime}$, or a linear size of 0.6 kpc (somewhat smaller than the \HI\ hole itself, which is 0.9 kpc in diameter from rim to rim). The position angle of the gradient is about $225^\circ$, similar to the local orientation of the ordered magnetic field as can be clearly seen in Figure \ref{fig:rmhole}. By measuring the typical fluctuations in the RM values on the same angular scale throughout the rest of the galaxy, this particular feature is found to be significant (i.e., distinguishable from a noise feature) at the $4\sigma$ level. Here, $\sigma$ includes both noise and contributions from real fluctuations in $\vec{B}$ and $n_e$ on 0.6 kpc scales throughout the star forming disk. Very few RM fluctuations with similar magnitude are found, and the others are preferentially at the edge of the polarized disk with low confidence. We can also assess the probability of a chance association of this resolved RM feature with an \HI\ hole. This is done by computing the density of \HI\ hole centers within the regions with well-detected diffuse polarized emission, taking into account the quoted positional uncertainties of $4\arcsec\times4\arcsec$ \citep{boomsma_2007}. In this way, we estimate that the chance of a spurious spatial association of the RM feature and the \HI\ hole is approximately 0.5\%.

The RM gradient is centered at the typical RM value in the surrounding disk, making it unlikely to be caused by fluctuations in $n_e$. The RM gradient is, on the other hand, naturally explained by a significant vertical deviation of the local ordered magnetic field direction -- the field orientation pointing toward the observer on one side, and away on the other. The observed RM pattern originates on the front side of the disk because turbulence in the midplane depolarizes radiation from the backside \citep[e.g.,][]{braun_etal_2010}. Since RM is defined to be a positive quantity for a LOS magnetic field pointing toward the observer, the observed pattern implies that the field is oriented toward the observer on the side marked `A' in Figure~\ref{fig:rmhole}, and away on side `B'. Thus the ordered magnetic field is directed inward (along the spiral pattern toward the center of the galaxy). This agrees with the conclusion drawn from the much larger-scale RM gradient already observed across the entire disk \citep{beck_2007}.

\section{Discussion}

\begin{figure*}
\centering
\includegraphics[width=0.95\textwidth]{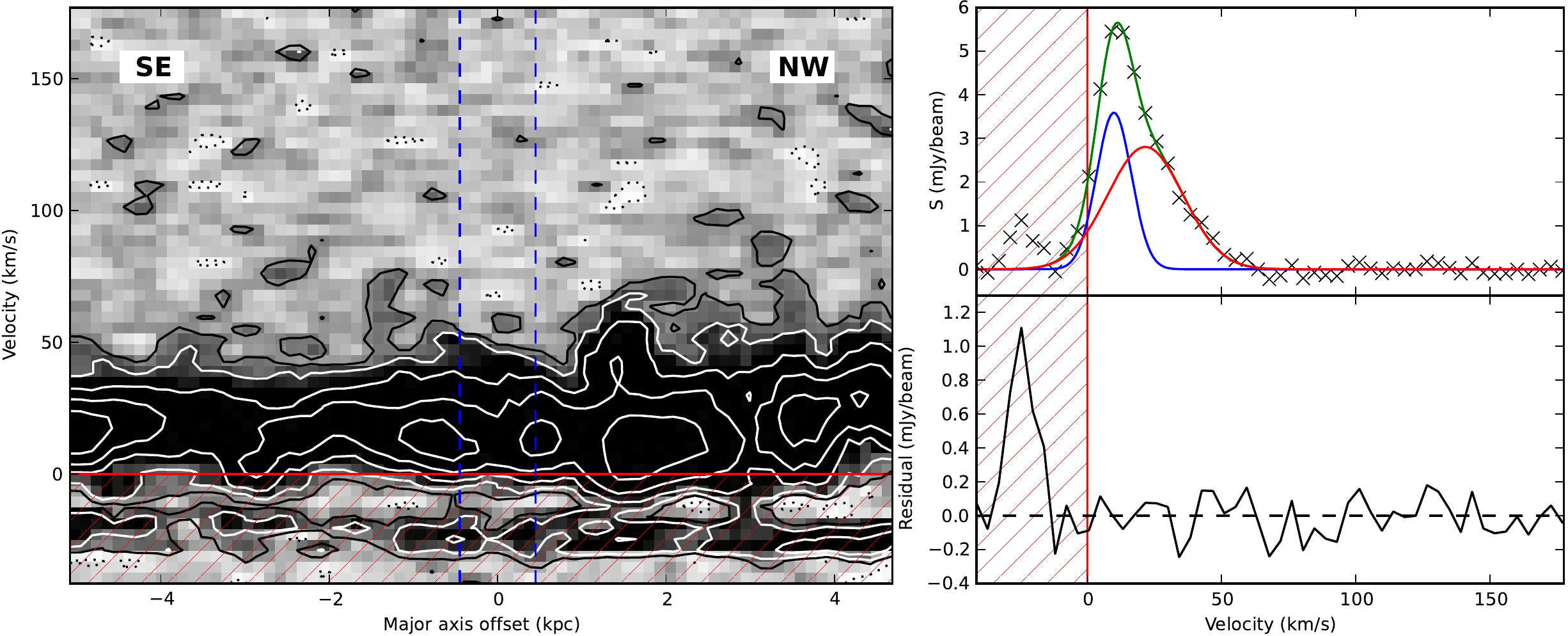}
\caption{Kinematics of the \HI\ hole. Left: PV diagram along the slice shown in the top left panel of Figure~\ref{fig:grid}. The orientation along the slice (SE to NW) is indicated. Contours start at $0.36\,\mathrm{mJy\,beam^{-1}}$ ($1.5\sigma$) and increase by powers of two. Dotted contours are drawn for negative values. The locations of the edge of the \HI\ hole are indicated with vertical dashed blue lines. Note the high-velocity wing at the location of the hole. The red hatched region (at negative velocities) shows the area of the diagram which is contaminated by MW \HI\ emission. Right: the average velocity profile within the hole region of the PV diagram, along with a two-component Gaussian fit to the profile. The red and blue components together give the combined fit, which is shown in green. The residuals of the fit are shown at the bottom. Again, the red-hatched region indicates the values that are contaminated by MW \HI\ (that region was not included in the fit).}
\label{fig:pv}
\end{figure*}

The coincidence of the RM gradient with the \HI\ hole strongly suggests that this magnetic structure is caused by the energetic process that led to the formation of the \HI\ hole itself.
The \HI\ properties of the hole indicate that it is a fairly young \citep[$\tau\sim20\,\mathrm{Myr}$;][]{boomsma_2007} but well-defined structure (albeit with low \HI\ column density contrast; see Figure~\ref{fig:grid}). A position-velocity (PV) diagram along a slice through the hole feature is presented in Figure~\ref{fig:pv}. The hole is classified by \citet{boomsma_2007} as Type 1 under the scheme described by \citet{brinks_bajaja_1986}, meaning that it is characterized by a gap in the PV diagram but without the appearance of an expanding bubble. This particular hole is most easily recognized in Figure~\ref{fig:pv} by inspection of the contours and noting the density enhancement at the edges of the hole. A two-component fit to the average velocity profile over the width of the hole is also shown. The combination of two Gaussian components produces an excellent match to the data. The velocity difference between the components is
$11.7\,\mathrm{km\,s^{-1}}$; correcting for the inclination angle of $38^\circ$, the velocity along the axis perpendicular to the disk is $14.8\,\mathrm{km\,s^{-1}}$. The velocity dispersions of the two components increases in the redshifted component to $14.1\,\mathrm{km\,s^{-1}}$ compared to $6.5\,\mathrm{km\,s^{-1}}$ in lower-velocity component, which we associate with the ISM in the main disk. Assuming that the redshifted gas is located on the front side along with the magnetic feature, the second \HI\ component represents gas which is falling back onto the disk at an average velocity of $15\,\mathrm{km\,s^{-1}}$. Inspection of the PV diagram shows that the infall is more pronounced on the SE (radially outward) side of the hole. Such wings in the velocity profiles are ubiquitous in NGC~6946 and are certainly not unique to this region, but this particular redshifted feature fits naturally into the picture described below. Note that a blueshifted component, if present, would be confused by \HI\ emission from the Milky Way as indicated in Figure~\ref{fig:pv}.

The magnetic field strength in the vicinity of the bubble can be estimated by taking advantage of the fact that the observed RM gradient ($\Delta\mathrm{RM}$) is symmetric about the average RM in the region. Half of the RM gradient can thereby be attributed to the field strength, which is to say that
\begin{equation}
\frac{1}{2}\,\Delta\mathrm{RM}\,=\,0.81\int\,n_e\,\vec{B}\cdot\mathrm{d}\vec{l}.
\end{equation}
Taking the inclination $i$ of the galaxy into account, we have
\begin{equation}
\frac{1}{2}\,\Delta\mathrm{RM}\,=\,0.81\,n_e\,|\vec{B}|\,L\,\sin\alpha\sin i,
\end{equation}
where $L$ is the path length through the volume of interest and $\alpha$ is the angle between the magnetic field lines and the disk plane. Taking the inclination to be $38^\circ$, and estimating $\alpha=45^\circ$, $L=300\,\mathrm{pc}$ (half of the distance across the RM feature), and a typical thermal electron density $n_e=0.05\,\mathrm{cm^{-3}}$ \citep[cf.][]{ferriere_2001}, the resulting regular magnetic field strength is determined to be
\begin{equation}
|\vec{B}|\,=\,7.2\left[\left(\frac{n_e}{0.05\,\mathrm{cm^{-3}}}\right)\left(\frac{L}{300\,\mathrm{pc}}\right)\left(\frac{\sin\alpha}{\sin45^\circ}\right)\right]^{-1}\,\mu\mathrm{G},
\end{equation}
which compares well with the magnetic field strength previously measured in the range $8-10\,\mu\mathrm{G}$ in the magnetic arms \citep{beck_2007}, as is the feature studied here.

Taken together, these lines of evidence provide new observational support to a picture consistent with the chimney model \citep{norman_ikeuchi_1989} of disk-halo interaction, in which star formation activity in the disk drives convective hot gas flows upward into the halo, carrying magnetic fields along with the hot gas motion, and forming bubbles in the disk that are observable as \HI\ holes. The hot gas cools in the halo, condensing into \HI\ clouds which then return to the disk on time scales of a few tens of Myr and with a net outward radial movement \citep{collins_etal_2002,fraternali_binney_2006}. State-of-the-art numerical models \citep[e.g.,][]{korpi_etal_1999,avillez_breitschwerdt_2005} depict the detailed evolution of the ISM during this process. The observational results presented here make the first clear identification of magnetic involvement in disk-halo flows. This has significant consequences for our understanding of the galactic dynamo process, because large-scale magnetic fields are carried upward by the hot gas, but will not return downward with the cool clouds traced by the redshifted \HI\ \citep{brandenburg_etal_1995}. Moreover, the smaller-scale turbulent component of the magnetic field will also be carried out of the disk, which has been proposed as a mechanism to prevent quenching of the dynamo process \citep{shukurov_etal_2006}. As a related point, we remark that a close connection in NGC~6946 between the position angle of the elliptically shaped \HI\ holes and the local spiral pitch angle has previously been noticed \citep{boomsma_etal_2008}, and may actually reflect the fact that superbubbles preferentially expand along the direction of the ordered field \citep{tomisaka_1998,stil_etal_2009}.

It is remarkable that of the many \HI\ holes in NGC~6946 \citep[121 are listed by][]{boomsma_etal_2008}, only one shows a clear RM signature. In estimating how unique this feature may be, there are several factors to take into consideration.
\begin{itemize}
\item The lack of polarized synchrotron radiation on the receding (southwest) side of the galaxy, clearly seen in Figure~\ref{fig:rmhole}, is now understood to be a geometric effect \citep{braun_etal_2010}, so no RM information is available for the subset of holes in that quadrant of the disk. This effect is relevant for about 30 (25\%) of the catalogued holes.
\item Many holes are located where more intense star formation is taking place, which increases the level of turbulence and thus the local degree of depolarization. This explains why there is an overall anticorrelation between the locations of \HI\ holes and regions with diffuse polarized synchrotron radiation. It is likely that the particular feature studied here is located some distance above (closer to the observer than) the turbulent midplane of NGC~6946, making it visible in diffuse polarization. Only 16 (13\%) of the catalogued \HI\ holes fall {\it fully} within the regions which do not suffer from depolarization effects or the geometric effect mentioned above (see Figure~\ref{fig:grid}). Estimating that the diffuse polarized emission originates at least one exponential scale height above the midplane on the frontside, only about 25\% of those 16 holes could have been traced by diffuse polarized emission.
\item The age of the hole must be within a certain range: old enough that the field configuration has had time to gain a significant vertical offset, but young enough that vertical shear has not yet destroyed the observational signature. For an outflow speed of $100\,\mathrm{km\,s^{-1}}$, the feature can grow to 300\,pc height after only about 3\,Myr. The vertical shear is not known for NGC~6946 \citep{boomsma_etal_2008}, but it is suspected to be at least as high as the measured value in NGC~891, $\approx15\,\mathrm{km\,s^{-1}\,kpc^{-1}}$ \citep{heald_etal_2006,oosterloo_etal_2007}. For this, the characteristic shear time would be around 60\,Myr. Other galaxies have higher shear values \citep[e.g.,][]{heald_etal_2007}, so the length of the observational window is short indeed but matches well with the age estimated for this particular hole from its \HI\ properties.
\item Finally, we note that the azimuthal location of the \HI\ hole within the disk of NGC~6946 is only $12^\circ$ from the minor axis. Since the pitch angle of the spiral pattern in the magnetic field vectors is about $18^\circ$ (cf. the optical spiral pitch angle given by \citet{kennicutt_1981}, $28^\circ$), this means that the location in the disk leads to the ideal geometrical situation for identifying an RM gradient across the feature. At other locations in the disk such a feature would have a different and far less obvious observational signature.
\end{itemize}
Considering all of these effects, we estimate that the feature described here could have been detected in only about 4 of the 121 catalogued \HI\ holes. It therefore seems reasonable to expect that the vertical magnetic field transport detected in this one \HI\ hole is actually much more ubiquitous across the disk of NGC~6946 and other similar galaxies. Sensitive observations at higher radio frequencies, where depolarization effects are less severe, can be utilized to test this prediction. Observations in the $2-4$ GHz range, for example, would provide a reasonable compromise between recovery of depolarized diffuse emission on the one hand, and acceptable RM precision on the other hand.

\acknowledgments

I thank Rense Boomsma and Tom Oosterloo for providing access to the \HI\ data and the list of \HI\ holes, as well as John McKean for helpful comments on an early version of the manuscript. I also thank the anonymous referee for comments that helped to strengthen the conclusions of the paper. The Second Palomar Observatory Sky Survey (POSS-II) was made by the California Institute of Technology with funds from the National Science Foundation, the National Geographic Society, the Sloan Foundation, the Samuel Oschin Foundation, and the Eastman Kodak Corporation. The Galaxy Evolution Explorer (GALEX) is a NASA Small Explorer. The mission was developed in cooperation with the Centre National d'Etudes Spatiales of France and the Korean Ministry of Science and Technology.

{\it Facilities:} \facility{WSRT}.

\end{document}